\documentclass{osa-article}

\journal{oe}

\articletype{Research Article}

\usepackage{changes}

\begin{document}

\title{Fiber-tip spintronic terahertz emitters}

\author{Felix Paries,\authormark{1,2,*} Nicolas Tiercelin,\authormark{3} Geoffrey Lezier,\authormark{3} Mathias Vanwolleghem,\authormark{3} Felix Selz,\authormark{1,2} Maria-Andromachi Syskaki,\authormark{4} Fabian Kammerbauer,\authormark{5} Gerhard Jakob,\authormark{5} Martin Jourdan,\authormark{5} Mathias Kläui,\authormark{5}  Zdenek Kaspar,\authormark{6} Tobias Kampfrath,\authormark{6} Tom S. Seifert,\authormark{6, 7}   Georg von Freymann,\authormark{1,2} and Daniel Molter\authormark{1}}

\address{\authormark{1}Fraunhofer Institute for Industrial Mathematics ITWM, Department Materials Characterization and Testing, 67663\,Kaiserslautern, GERMANY\\
\authormark{2}Department of Physics and Research Center OPTIMAS, RPTU Kaiserslautern-Landau, 67663\,Kaiserslautern, GERMANY\\
\authormark{3}University of Lille, CNRS, Centrale Lille, Université Polytechnique Hauts-de-France, UMR 8520 - IEMN, 59000\,Lille, FRANCE\\
\authormark{4}Singulus Technologies AG, Hanauer Landstrasse 107, 63796\,Kahl am Main, GERMANY\\
\authormark{5}Institute of Physics, Johannes-Gutenberg-Universität Mainz, 55128\,Mainz, GERMANY\\
\authormark{6}Department of Physics, Freie Universität Berlin, 14195\,Berlin, GERMANY\\ 
\authormark{7}TeraSpinTec GmbH, Lüneburger Str. 26, 10557 Berlin, GERMANY\\
\authormark{*}\email{felix.paries@itwm.fraunhofer.de} 
}



\begin{abstract}
Spintronic terahertz emitters promise terahertz sources with an unmatched broad frequency bandwidth that are easy to fabricate and operate, and therefore easy to scale at low cost. However, current experiments and proofs of concept rely on free-space ultrafast pump lasers and rather complex benchtop setups. This contrasts with the requirements of widespread industrial applications, where robust, compact, and safe designs are needed. To meet these requirements, we present a novel fiber-tip spintronic terahertz emitter solution that allows spintronic terahertz systems to be fully fiber-coupled. Using single-mode fiber waveguiding, the newly developed solution naturally leads to a simple and straightforward terahertz near-field imaging system with a 90\%-10\% knife-edge-response spatial resolution of 30\,µm.
\end{abstract}

\section{Introduction}
Spintronic terahertz emitters (STEs) convert an ultrashort laser pulse into an ultrabroadband terahertz pulse through a sequence of spin-current generation, spin-to-charge-current conversion, and current-to-field conversion within a thin-film stack of ferromagnetic metal (FM) and normal metal (NM) layers \cite{Seifert2022, Wu2021}. In 2010, M. Battiato, K. Carva, and P. M. Oppeneer theoretically predicted a spin-dependent superdiffusive transport of laser-excited electrons resulting in a strong spin injection from an FM into an NM layer \cite{Battiato2010}. In the following years, A. Melnikov \textit{et al.} were the first to confirm this effect experimentally \cite{Melnikov2011} and T. Kampfrath \textit{et al.} could show that this effect combined with the inverse spin Hall effect (ISHE) in an FM-NM heterostructure can lead to the emission of terahertz pulses. \cite{Kampfrath2013}. Latest since T. Seifert \textit{et al.} demonstrated an efficient tri-layer spintronic terahertz emitter in 2016, the field has attracted considerable interest \cite{Seifert2016}. It raised the prospect of terahertz sources with unmatched broad frequency bandwidth that could be scaled conveniently and cost effectively. Therefore, a considerable amount of research has been done to further improve the optical-to-terahertz conversion efficiency and the total terahertz output power. Detailed overviews can be found in the references \cite{Seifert2022, Bull2021, Feng2021, Beigang2020}. To highlight some examples, it has been shown that a (W/Fe/Pt/SiO$_{2}$)$_{\text{2}}$ tri-layer stack can yield a 1.7 fold increase in terahertz conversion efficiency compared to a single (W/Fe/Pt/SiO$_{2}$)$_{\text{1}}$ tri-layer structure \cite{Feng2018}. Similarly, a photonic cavity structure was shown to increase the terahertz conversion efficiency by a factor of 1.7 \cite{Kolejak2022}. In addition to improving the laser pumping process, it was shown that attaching a hyper-hemispherical silicon lens \cite{Torosyan2018} or coupling an H-dipole antenna \cite{Nandi2019} to the STE significantly improves the terahertz outcoupling process. In addition, groundbreaking proofs of concept for a variety of possible applications have been demonstrated: Since the emitted terahertz electric field is to a very good approximation perpendicular to the FM magnetization \cite{Seifert2022}, controlling and modulating the direction of the FM magnetization promises applications such as spintronic-based terahertz ellipsometry or terahertz wireless communication. The FM magnetization can be controlled not only by rotating the external magnetic field \cite{Gueckstock2021, Kong2019}, but also by a simple bipolar variation of the magnetic field strength \cite{Kolejak2022_ACS}, or by magneto-electrically inducing a mechanical strain \cite{Lezier2022}. In addition to these achievements in terahertz polarization control, a terahertz magneto-optical imager consisting of an electro-optic crystal sensor stacked on an STE is noteworthy \cite{Bulgarevich2020}.

However, in these demonstrations, free-space ultrashort-pulsed pump lasers are used alongside rather complex benchtop setups. This does not meet industrial requirements for robustness, compactness, and safety. A transition from free-space laboratory setups to fully fiber-based systems is therefore desirable.

Here we present a novel fiber-tip spintronic terahertz emitter solution that allows spintronic terahertz systems to be fully fiber-coupled. This is an important step towards industrial-grade systems that exploit the many advantages of STEs.

As will be shown in the following sections, our novel fiber-tip spintronic terahertz emitter solution not only makes full fiber-coupling possible, but also inherently provides terahertz sources with sub-wavelength dimensions that are suitable for broadband terahertz near-field imaging \cite{Adam2011, Mittleman2018}. In comparison to the fiber-based terahertz near-field imaging sources developed by M. Yi \textit{et al.} who created a terahertz source with sub-wavelength dimensions by bonding an epitaxially grown InAs crystal to a 45-degree wedge end facet of an optical fiber \cite{Yi2010}, our terahertz near-field imaging sources require a less complex fabrication process, can be operated wavelength-independently, and exhibit higher bandwidth as well as higher near-field resolution. Integrating one of our fiber-tip STEs into a standard terahertz time-domain spectroscopy setup seamlessly leads to a terahertz near-field imaging system, that is much simpler and more robust than state-of-the-art terahertz scattering-type scanning near-field optical microscopy (THz s-SNOM) systems \cite{Keilmann2004, Huber2008, Eisele2014, Murakami2014, Giordano2018, Thomas2022}. As a final remark, two other noteworthy spintronic-based concepts for terahertz near-field imaging were published recently: Z. Bai \textit{et al.} demonstrated a spintronic terahertz near-field biosensing approach \cite{Bai2020} and Chen \textit{et al.} presented a spintronic microscopy concept based on the ghost near-field imaging technique with sub-10\,µm spatial resolution \cite{Chen2020}. While the first is very limited in bandwidth and tailored toward the demonstrated use case, the latter requires again a complex laboratory setup including free-space lasers and two digital micromirror devices as programmable masks making these approaches cumbersome.
\newpage
\section{Fiber-tip spintronic terahertz emitters}
\begin{figure}[h!]
    \centering\includegraphics[width=0.9\textwidth]{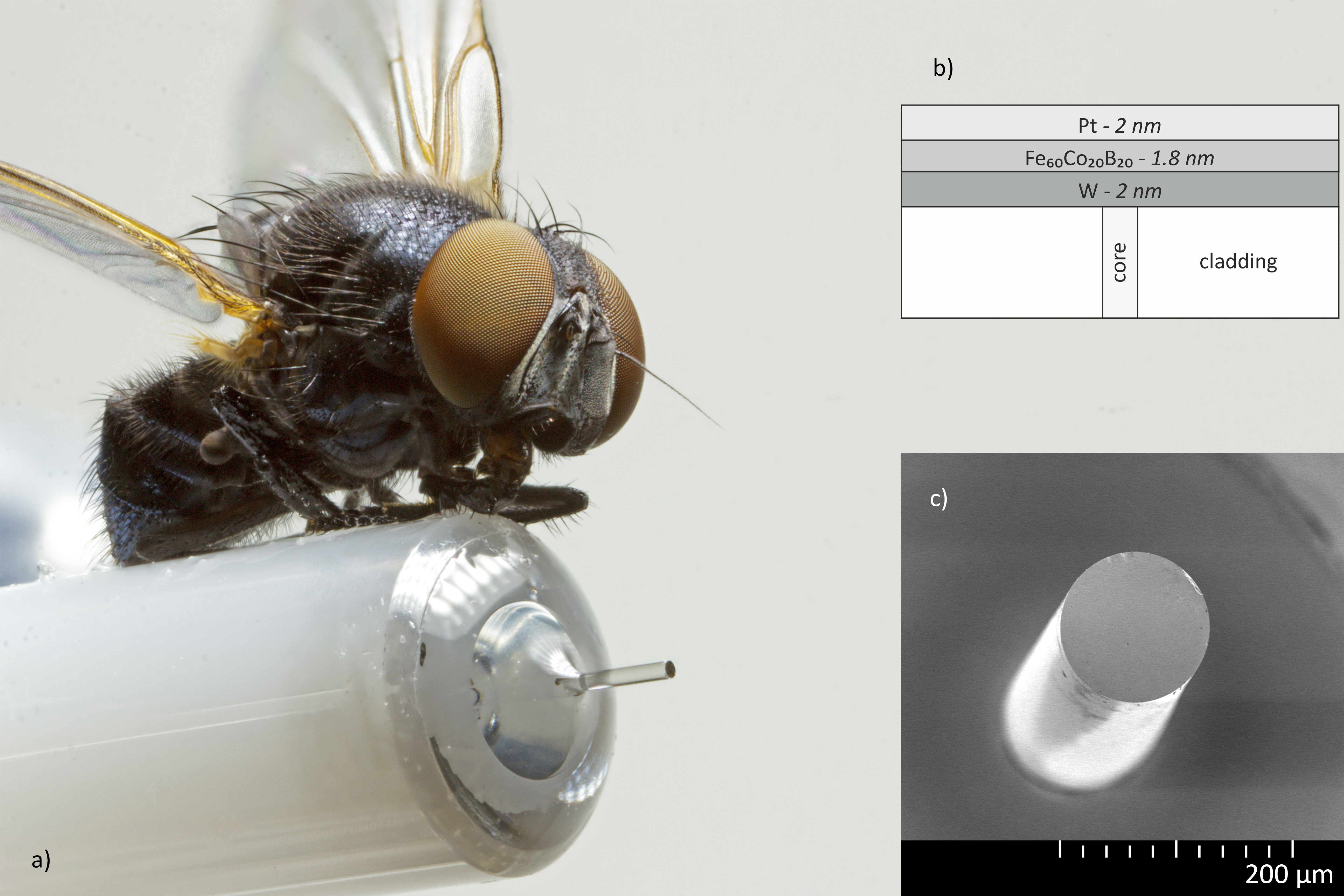}
    \caption{a) Fiber-tip spintronic terahertz emitter. For convenient handling and integration, as well as easy pump-fiber coupling via an FC/PC connector, the fiber-tip STE is enclosed in a standard 2.5\,mm fiber ferrule. A \textit{Calyptratae} is sitting on top for intuitive scale comparison. These fiber-tip spintronic terahertz emitters can be used for the transition from a free-space setup to a fully fiber-based system. b) Sketch of the spintronic tri-layer structure used on a single-mode glass fiber with a typical step-index profile dividing the waveguide into core and cladding. c) Scanning electron microscope image of the fiber-tip STE shown in a). The outer diameter is 125\,µm, while the mode field diameter of the pump field is only about 10\,µm due to the single-mode waveguiding within this fiber.}
    \label{fig:STE}
\end{figure}

In our new concept, a spintronic tri-layer structure [W(2.0 nm)\,/\,FeCoB(1.8 nm)\,/\,Pt(2.0 nm)] is sputtered directly onto the tip of a fiber. Therefore, we call it a fiber-tip spintronic terahertz emitter. Fig.~\ref{fig:STE} shows an example of such a fiber-tip STE. The fiber is enclosed in a standard 2.5\,mm fiber ferrule to achieve two advantages: first, it makes handling and integration very convenient, and second, the pump fiber can be connected via a standard fiber-optic connector with physical contact (FC/PC). The scanning electron microscopy zoom shown in Fig.~\ref{fig:STE}\,c) reveals the surface and dimensions of the fiber-tip STE from Fig.~\ref{fig:STE}\,a). Using single-mode fibers optimized for our pump wavelength of 1550\,nm, the pump mode field diameter at the surface can be reduced to only about 10\,µm. We have fabricated fiber-tip STEs with two different fiber materials (glass and sapphire) and with different diameters ranging from 100\,µm to 425\,µm. 

\section{Methods}
\subsection{Setup}
To characterize the fiber-tip spintronic terahertz emitters, we used a standard fiber-based terahertz time-domain spectroscopy setup consisting of a single mode-locked Erbium-doped fiber laser (ML EDFL) with two equal output ports, each emitting a pulse train with 100\, MHz repetition rate, 1560\,nm center wavelength, and 70\,fs FWHM temporal pulse width at the emitter and detector positions, two identical Erbium-doped fiber amplifiers (EDFA), and a delay stage. The pump fiber was butt-coupled to the fiber-tip STE via a simple FC/PC connection with reflection losses of less than 5\,\%. The magnetic field was provided by two small NdFeB magnets integrated near the fiber tip in a 3D printed one\,inch ferrule holder. The emitted terahertz signal was modulated by a rotating chopper blade and focused by two parabolic mirrors onto a photoconductive antenna (PCA) detector. The induced current was converted to voltage by a trans-impedance amplifier and finally detected by a lock-in amplifier. For terahertz near-field imaging, a high-resolution metal mask was mounted on a motorized XYZ stage and moved in front of the fiber-tip STE. 

\begin{figure}[h!]
    \centering\includegraphics[width=0.9\textwidth]{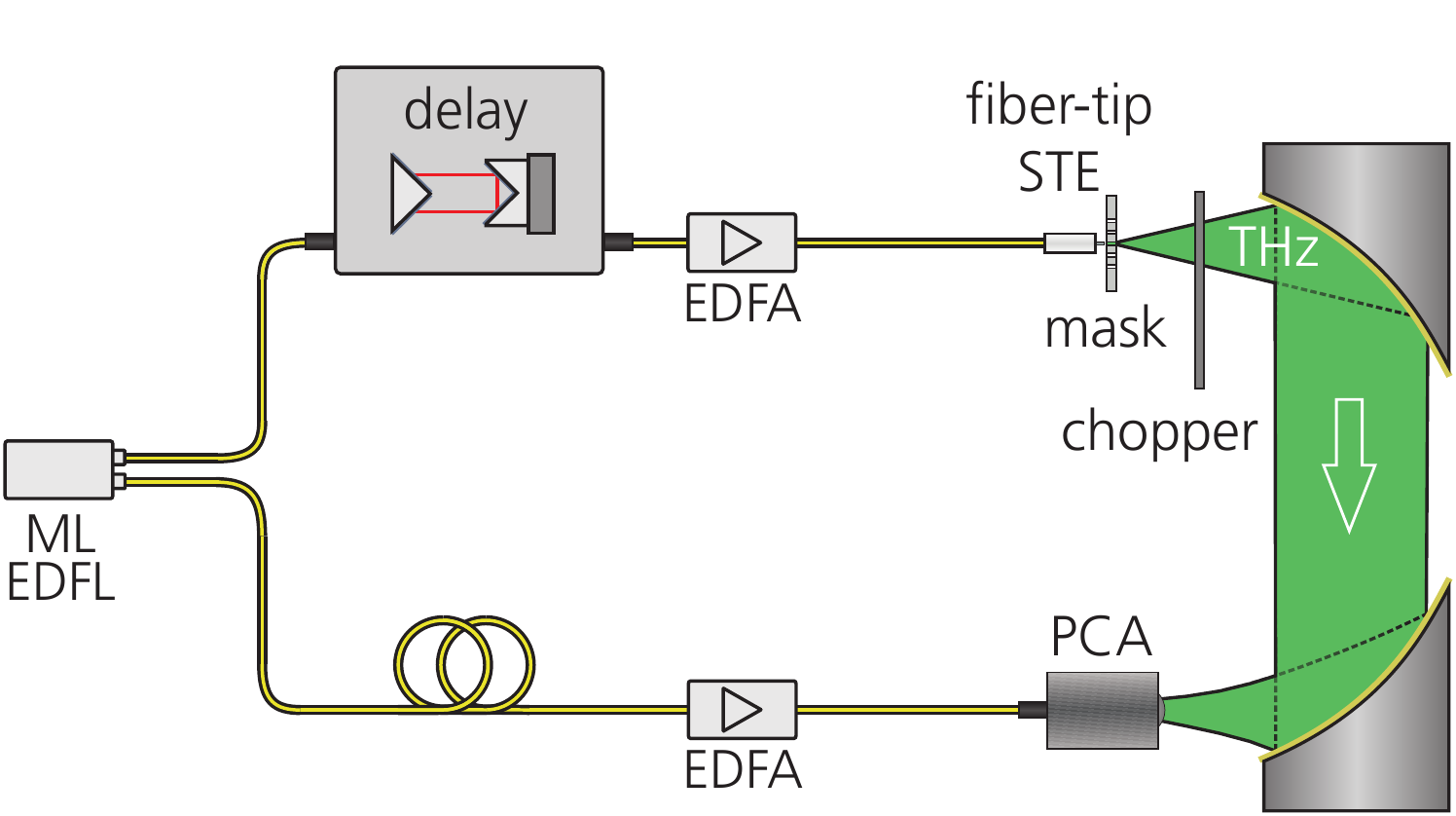}
    \caption{Experimental setup. A standard fiber-based terahertz time-domain spectroscopy setup has been used. A mode-locked Erbium-doped fiber laser (ML EDFL) emits two equal pulse trains which are amplified using Erbium-doped fiber amplifiers (EDFA). For the pump-probe experiment to be conducted, a delay line adjusts the two pulse trains' time delay. The pump fiber is butt-coupled to the fiber-tip STE via a simple FC/PC connection and the emitted terahertz radiation is focused onto a photo-conductive antenna (PCA) detector. A movable high-resolution mask allows for high-resolution terahertz near-field imaging.}
    \label{fig:setup}
\end{figure}

\subsection{Fabrication}
The spintronic W(2.0 nm)\,/\,FeCoB(1.8 nm)\,/\,Pt(2.0 nm) tri-layer structure was deposited on the fiber tips using a thin-film sputtering process. Prior to deposition, a cleaning procedure involved 5\,min ultrasonication steps in successive baths of acetone, isopropyl alcohol, and de-ionized water. The W/FeCoB/Pt spintronic emitters were deposited by RF-diode sputtering in LEYBOLD Z550 equipment where the individual layers are sputtered from circular 4 inches targets. In particular, the Fe$_{60}$Co$_{20}$B$_{20}$ alloy is obtained from a target with this stoichiometry. To ensure high precision in the thicknesses, the deposition was carried on a rotatory turn-table substrate holder in an oscillation mode: each time the substrate passes under the target, which corresponds to one oscillation, a layer of the material is deposited with a thickness that depends on the angular speed of the turn-table. The deposition is carefully calibrated beforehand by typically depositing a film with 100\,oscillations.

For the W(2.0\,nm)\,/\,FeCo(0.5\,nm)/TbCo$_{2}$(0.8\,nm)/FeCo(0.5\,nm)\,/\,Pt(2.0\,nm) emitters, the TbCo$_2$ and FeCo alloys composing the ferromagnetic layer are obtained from composite targets. Given their thickness, the CoFe/TbCo$_2$/CoFe tri-layer acts as an exchange-coupled multilayer. During deposition, the tips are placed in an in-plane field with an approximate strength of 80\,kA/m to imprint an in-plane magnetic anisotropy along a chosen direction in the ferromagnetic tri-layer. 
\newpage
\section{Results}
\subsection{Terahertz Signals}
\begin{figure}[h!]
    \centering\includegraphics[width=0.99\textwidth]
    {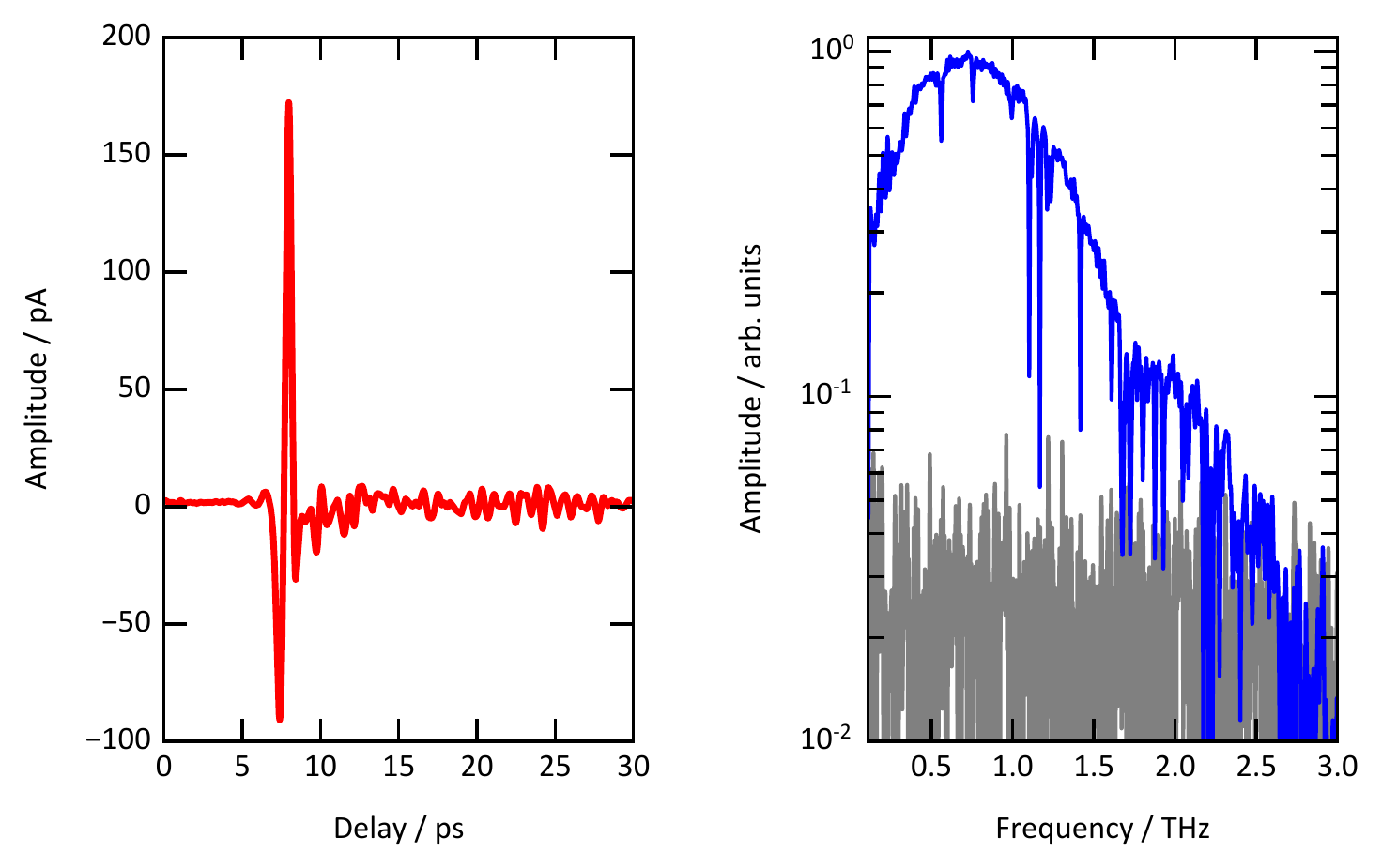}
    \caption{Terahertz signal (left) and frequency spectrum with noise level (right) of a fiber-tip spintronic terahertz emitter. The ultrashort-pulse-pumped fiber-tip spintronic terahertz emitter emits a typical terahertz waveform and the spectrum reveals the characteristic water vapor absorption lines. This proves the full functionality of our newly developed concept.}
    \label{fig:THzSignal}
\end{figure}

Figure~\ref{fig:THzSignal} exemplary shows a 30\,ps section of a recorded terahertz waveform with a time-domain window of 100\,ps and the associated frequency spectrum. We achieved a dynamic range of 28\,dB and a bandwidth of 2.5\,THz at an average pump power of 25\,mW. This proves the full functionality of our newly developed concept. It is worth noting that the simple FC/PC butt coupling of the pump fiber allows us to exchange and operate a fiber-tip spintronic emitter within a few minutes.  

\subsection{Near-field imaging}
\begin{figure}[h!]
    \centering\includegraphics[width=0.99\textwidth]{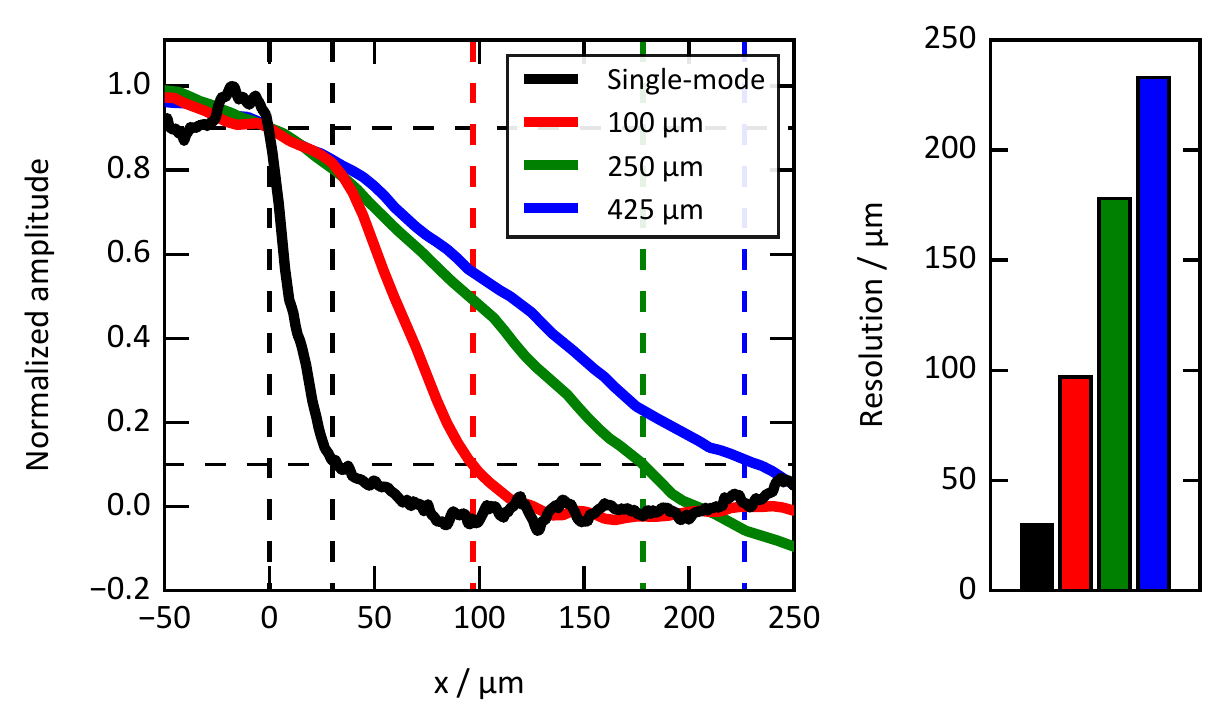}
    \caption{Knife-edge responses of four different fiber-tip spintronic terahertz emitters varying in fiber type and diameter. The coreless sapphire fibers with a diameter of 425\,µm, 250\,µm, and 100\,µm result in a 90\%-10\% knife-edge resolution of 233\,µm, 178\,µm, and 97\,µm, respectively. With the single-mode glass fiber a 90\%-10\% knife-edge resolution of 30\,µm could be achieved.}
    \label{fig:KnifeEdge}
\end{figure}

\begin{figure}[h!]
    \centering\includegraphics[width=0.99\textwidth]{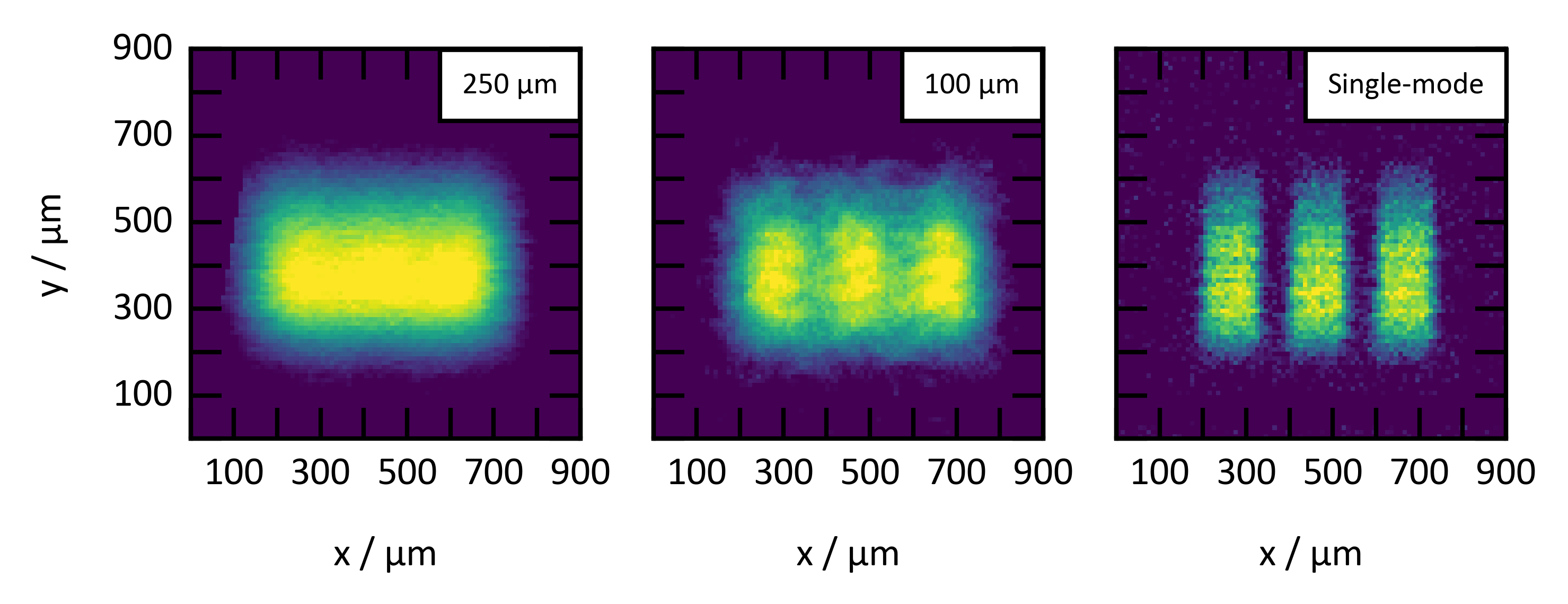}
    \caption{Near-field imaging of two thin metal stripes with a width of approximately 77.5\,µm and a spacing of approximately 122.5\,µm. While the metal stripes cannot be resolved using the 250\,µm fiber-tip spintronic terahertz emitter and only hinted at with the 100\,µm emitter, the single-mode version reveals the two metal stripes clearly. }
    \label{fig:NearFieldImaging}
\end{figure}

 Figure~\ref{fig:KnifeEdge} shows the knife-edge responses for four different fiber-tip STEs varying in fiber type and diameter: a standard single-mode glass fiber and three coreless sapphire fibers with outer diameters of 100\,µm, 250\,µm, and 425\,µm, respectively. The measurements confirm the expected mode field diameter dependent near field resolution. Using the single-mode fiber tip STE with a pump mode field diameter of approximately 10.5\,µm (FWHM), we achieved a 90\%-10\% knife-edge resolution of 30\,µm. The coreless sapphire fibers with diameters of 425\,µm, 250\,µm, and 100\,µm resulted in 90\%-10\% knife-edge resolutions of 233\,µm, 178\,µm, and 97\,µm, respectively. To test these capabilities, we performed 2D near-field imaging of two metal strips for the different emitter types with the maximum of the terahertz waveform as the working point. The metal strips have a width of about 77.5\,µm and a spacing of about 122.5\,µm (optical microscopy image readout). As can be seen in Fig.~\ref{fig:NearFieldImaging}, the stripes are clearly resolved with the single-mode fiber-tip spintronic emitter, only hinted at with the 100\,µm emitters, and not resolved with the 250\,µm emitter. 

\section{Challenges and outlook}
\subsection{Magnetic-bias-free fiber-tip STE}
\begin{figure}[h!]
    \centering\includegraphics[width=0.99\textwidth]{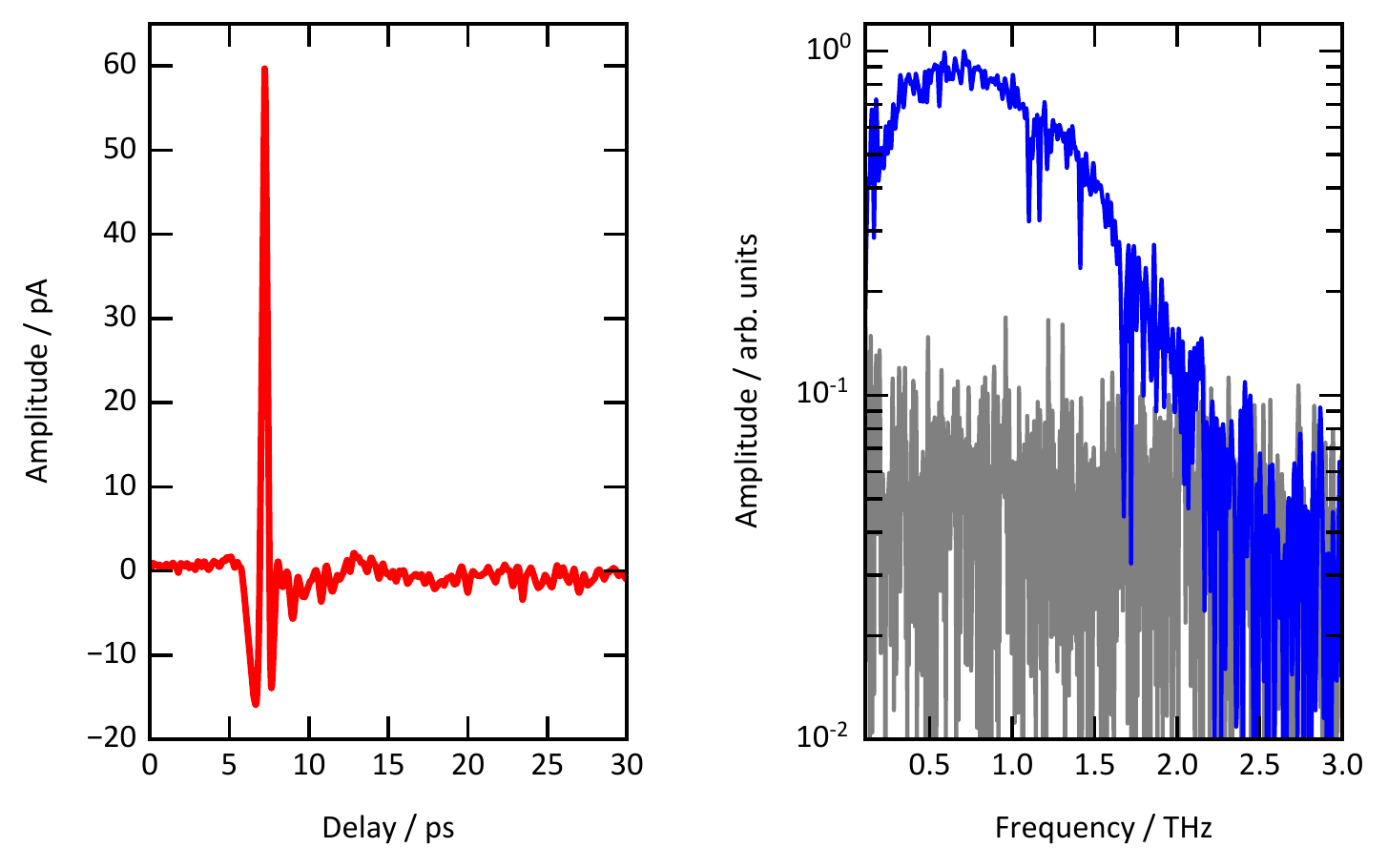}
    \caption{Terahertz signal (left) and frequency spectrum with noise level (right) from a magnetic-bias-free fiber-tip STE: the ferromagnetic FeCoB layer has been replaced with a hard-magnetic FeCo/TbCo$_{2}$/FeCo anisotropic heterostructure. This allows the fiber-tip STE to be operated without any external magnetic bias, as the heterostructure features a large coercive field and a sizeable remanent magnetization. Even though the peak-to-peak amplitude is lower than for the FeCoB fiber-tip STEs with external magnetic bias under the same conditions, the magnetic-bias-free fiber-tip STE is fully functional and the water vapor absorption lines are clearly visible.}
    \label{fig:HardMagnetic}
\end{figure}

The results presented above were achieved using the well-known W/FeCoB/Pt thin film design. However, in this design, the generation of terahertz pulses still relies on an external magnetic bias, which could become a challenge for small devices and lab-on-a-chip technologies. Replacing the ferromagnetic FeCoB layer with an anisotropic FeCo/TbCo$_{2}$/FeCo heterostructure solves this challenge and allows the fiber-tip STEs to be operated without an external magnetic bias \cite{Kolejak2022_ACS}. Figure~\ref{fig:HardMagnetic} shows the recorded terahertz signal under the same conditions as Fig.~\ref{fig:THzSignal}. Although the peak-to-peak amplitude is lower than that of the FeCoB fiber-tip STEs with external magnetic bias, we consider a dynamic range of 21.5\,dB a decent performance and an interesting opportunity to explore further. A challenge to be addressed is the steady demagnetization due to heat accumulation at higher pump powers. 

\subsection{Material destruction}
\begin{figure}[h!]
    \centering\includegraphics[width=0.99\textwidth]{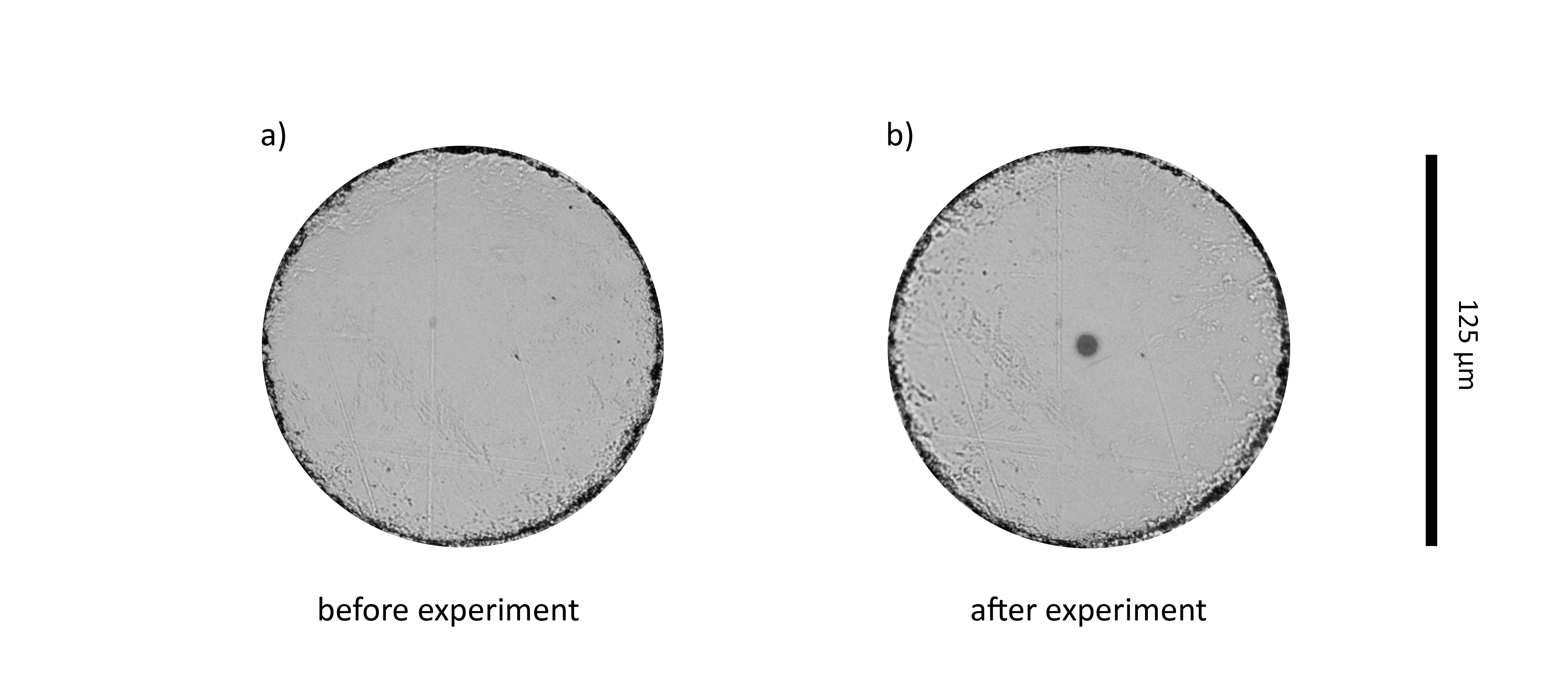}
    \caption{Material destruction of a single-mode glass-fiber-tip STE. While pumping above the pump fluence threshold, the material is steadily darkening within the area of the pump mode field. A visual comparison with the material before the experiment (a) reveals this phenomenon clearly (b).}
    \label{fig:MaterialDestruction_Image}
\end{figure}

\begin{figure}[h!]
    \centering\includegraphics[width=0.99\textwidth]{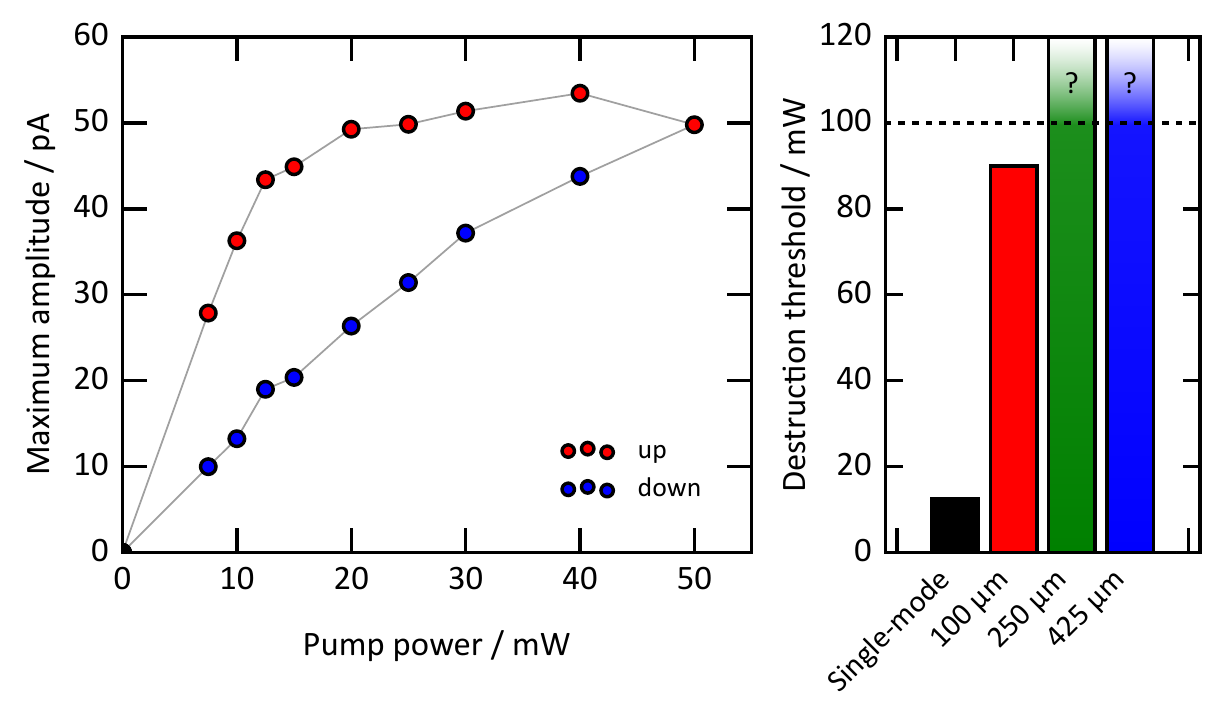}
    \caption{Material destruction hysteresis of a single-mode glass-fiber-tip STE (left) and pump-power destruction thresholds for various fiber-tip STEs (right). An increasing pump power causes a steady decrease in terahertz conversion efficiency once a specific threshold is reached. Decreasing the pump power again proves this change to be irreversible. The bar chart on the right shows the destruction threshold for various fiber-tip STEs and reveals the pump-mode-field-diameter dependence: The pump-power destruction threshold of a single-mode glass-fiber-tip STE is approximately 12\,mW, while the 100\,µm sapphire-fiber-tip STE exhibits a threshold of approximately 90\,mW. The 250\,µm und 425\,µm fiber-tip STEs were stable and did not show any material destruction at a pump power of 100\,mW. Due to the available fiber-laser setup, the pump power could not be increased above 100\,mW.}
    \label{fig:MaterialDestruction_Graph}
\end{figure}

A second challenge is a material change that we observe for pump powers above a mode field diameter dependent threshold. Since the terahertz signal decreases steadily on a second to minute time scale once the pump threshold is reached, and since we observe a darkening of the material in the region of the pump field (Fig.~\ref{fig:MaterialDestruction_Image}), we are confident that this is an irreversible chemical process due to the accumulation of heat. The fact that the material destruction threshold scales with increasing diameter supports this hypothesis (Fig.~\ref{fig:MaterialDestruction_Graph}). Approximating the fluences based on the knife-edge responses shown in Fig.~\ref{fig:KnifeEdge} and assuming a uniform pump mode field distribution, the fluence threshold is below 2\,kW/cm$^2$ (0.02\,mW/µm$^2$) for both sapphire and glass fiber tip STEs.

\newpage
\section{Conclusions}
In conclusion, we introduced our newly developed fiber-tip spintronic terahertz emitter solution, which allows spintronic terahertz systems to be fully fiber-coupled. The pump fiber can be connected via a standard FC/PC connector, and the emitters can be easily changed within a few minutes. We have recorded terahertz signals to demonstrate full functionality, and our near-field imaging with a 90\%-10\% knife-edge resolution of 30 µm demonstrates a possible application. In addition, we presented an advanced magnetic-bias-free fiber-tip spintronic terahertz emitter and addressed the challenge of an existing material destruction threshold limiting the pump power. We anticipate our solution to lay the groundwork for a far-reaching transition from free-space setups to fiber-coupled systems and to catalyze further research in the field of fiber-tip spintronic terahertz emitters.

\section{Acknowledgement}
The authors gratefully acknowledge funding from the Horizon 2020 European Union Research and Innovation program under FET-OPEN Grant agreement No. 863155 (S-Nebula) and Marie Skłodowska-Curie Grant agreement No. 860060 “Magnetism and the effect of Electric Field” (MagnEFi), as well as from the Deutsche Forschungsgemeinschaft (DFG, German Research Foundation) - TRR 173 - 268565370 (SPIN+X, projects A01, B02, and B11). Moreover, the authors would like to thank the French RENATECH Network and the Nano Structuring Center at the RPTU Kaiserslautern-Landau for their technical support during the development of the fiber-tip spintronic terahertz emitters.

\section{Disclosures}
The authors declare no conflicts of interest.

\section{Data availability}
The data that support the findings of this study are openly available at \newline https://doi.org/10.5281/zenodo.7885846 .

 

\begin{thebibliography}{1}
 \newcommand{\enquote}[1]{``#1''}

\bibitem{Seifert2022}
T. Seifert, L. Cheng, Z. Wei, T. Kampfrath, and J. Qi \enquote{Spintronic sources of ultrashort terahertz electromagnetic pulses}, Appl. Phys. Lett. \textbf{120} (2022).

\bibitem{Wu2021}
W. Wu, C.Y. Ameyaw, M. Doty, and M. B. Jungfleisch \enquote{Principles of spintronic THz emitters}, J. Appl. Phys. \textbf{130} (2021).

\bibitem{Battiato2010}
M. Battiato, K. Carva, and P.M. Oppeneer \enquote{Superdiffusive Spin Transport as a Mechanism of Ultrafast Demagnetization}, Phys. Rev. Let. \textbf{105} (2010).

\bibitem{Melnikov2011}
A. Melnikov, I. Razdolski, T. O. Wehling, E. Th. Papaioannou, V. Roddatis, P. Fumagalli, O. Aktsipetrov, A. I. Lichtenstein, and U. Bovensiepen \enquote{Ultrafast Transport of Laser-Excited Spin-Polarized Carriers in Au/Fe/MgO(001)}, Phys. Rev. Let. \textbf{107} (2011).
 
\bibitem{Kampfrath2013}
T. Kampfrath, M. Battiato, P. Maldonado, G. Eilers, J. Nötzold, S. Mährlein, V. Zbarsky, F. Freimuth, Y. Mokrousov, S. Blügel, M. Wolf, I. Radu, P.M. Oppeneer, and M. Münzenberg \enquote{Terahertz spin current pulses controlled by magnetic heterostructures}, Nature Nanotechnology \textbf{8} (2013).

\bibitem{Seifert2016}
T. Seifert, S. Jaiswal, U. Martens, J. Hannegan, L. Braun, P. Maldonado, F. Freimuth, A. Kronenberg, J. Henrizi, I. Radu, E. Beaurepaire, Y. Mokrousov, P.M. Oppeneer, M. Jourdan, G. Jakob, D. Turchinovich, L.M. Hayden, M. Wolf, M. Münzenberg, M. Kläui, and T. Kampfrath, \enquote{Efficient metallic spintronic emitters of ultrabroadband terahertz radiation}, Nature Photonics \textbf{10} (2016).

\bibitem{Bull2021}
C. Bull, S.M. Hewett, R. Ji, C.-H. Lin, T. Thomson, D.M. Graham, and P. W. Nutter \enquote{Spintronic terahertz emitters: Status and prospects from a material perspective}, APL Mater. \textbf{9} (2021).

\bibitem{Feng2021}
Z. Feng, H. Qiu, D. Wang, C. Zhang, S. Sun, B. Jin, and W. Tan \enquote{Spintronic terahertz emitter}, J. Appl. Phys. \textbf{129} (2021).

\bibitem{Beigang2020}
E.T. Papaioannou, and R. Beigang \enquote{THz spintronic emitters: a review on achievements and future challenges}, De Gruyter Nanophotonics \textbf{10} (2021).

\bibitem{Feng2018}
Z. Feng, R. Yu, Y. Zhou, H. Lu, W. Tan, H. Deng, Q. Liu, Z. Zhai, L. Zhu, J. Cai, B. Miao, and H. Ding \enquote{Highly Efficient Spintronic Terahertz Emitter Enabled by Metal–Dielectric Photonic Crystal}, Adv. Optical Mater. \textbf{6} (2018).

\bibitem{Kolejak2022}
P. Koleják, G. Lezier, L. Halagačka, Z. Gelnárová, J.-F. Lampin, N. Tiercelin, M. Vanwolleghem, and K. Postava  \enquote{Highly efficient terahertz spintronic emitter integrated with optimized photonic crystal}, 47th IRMMW-THz (2022).

\bibitem{Torosyan2018}
G. Torosyan, S. Keller, L. Scheuer, R. Beigang, and E.Th. Papaioannou \enquote{Optimized Spintronic Terahertz Emitters Based on Epitaxial Grown FE/PT Layer Structures}, Scientific Reports \textbf{8} (2018).

\bibitem{Nandi2019}
U. Nandi, M.S. Abdelaziz, S. Jaiswal, G. Jakob, O. Gueckstock, S.M. Rouzegar, T.S. Seifert, M. Kläui, T. Kampfrath, S. Preu \enquote{Antenna-coupled spintronic terahertz emitters driven by a 1550 nm femtosecond oscillator}, Appl. Phys. Lett. \textbf{115} (2019).

\bibitem{Gueckstock2021}
O. Gueckstock, L. Nádvorník, T.S. Seifert, M. Borchert, G. Jakob, G. Schmidt, G. Woltersdorf, M. Kläui, M. Wolf, and T. Kampfrath \enquote{Modulating the polarization of broadband terahertz pulses from a spintronic emitter at rates up to 10 kHz} Optica Vol. 8, No. 7 (2021).

\bibitem{Kong2019}
D. Kong, X. Wu, B. Wang, T. Nie, M. Xiao, C. Pandey, Y. Gao, L. Wen, W. Zhao, C. Ruan, J. Miao, Y. Li, and L. Wang \enquote{Broadband Spintronic Terahertz Emitter with Magnetic-Field Manipulated Polarizations} Adv. Optical Mater. \textbf{7} (2019).

\bibitem{Kolejak2022_ACS}
P. Koleják, G. Lezier, K. Postava, J.-F. Lampin, N. Tiercelin, and M. Vanwolleghem \enquote{360° Polarization Control of Terahertz Spintronic Emitters Using Uniaxial FeCo/TbCo2/FeCo Trilayers} ACS Photonics \textbf{9} (2022).

\bibitem{Lezier2022}
 G. Lezier,  P. Koleják,  J.-F. Lampin,  K. Postava,  M. Vanwolleghem, and  N. Tiercelin \enquote{Fully reversible magnetoelectric voltage controlled THz polarization rotation in magnetostrictive spintronic emitters on PMN-PT} Appl. Phys. Lett. \textbf{120} (2022).

\bibitem{Bulgarevich2020}
D.S. Bulgarevich, Y. Akamine, M. Talara, V. Mag-usara, H. Kitahara, H. Kato, M. Shiihara, M. Tani and M. Watanabe  \enquote{Terahertz Magneto-Optic Sensor/Imager} Scientific reports \textbf{10} (2020).

\bibitem{Adam2011}
A.J.L. Adam \enquote{Review of Near-Field Terahertz Measurement Methods and Their Applications} J Infrared Milli Terahz Waves (2011).

\bibitem{Mittleman2018}
A.J.L. Adam \enquote{Twenty years of terahertz imaging} Optics Express \textbf{26} (2021).

\bibitem{Yi2010}
M. Yi, K. Lee, J. Lim, Y. Hong, Y.-D. Jho, and J. Ahn \enquote{Terahertz Waves Emitted from an Optical Fiber} Optics Express \textbf{18} (2010).

\bibitem{Keilmann2004}
F. Keilmann and R. Hillenbrand \enquote{Near-field microscopy by elastic light scattering from a tip} Phil. Trans. R. Soc. Lond. A \textbf{362} (2004).

\bibitem{Huber2008}
A.J. Huber, F. Keilmann, J. Wittborn, J. Aizpurua, and R. Hillenbrand \enquote{Terahertz Near-Field Nanoscopy of Mobile Carriers in Single Semiconductor Nanodevices} Nano Letters \textbf{8} (2008).

\bibitem{Eisele2014}
M. Eisele, T. L. Cocker, M. A. Huber, M. Plankl, L. Viti, D. Ercolani, L. Sorba, M. S. Vitiello, and R. Huber \enquote{Ultrafast multi-terahertz nano-spectroscopy with sub-cycle temporal resolution} Nature Photonics \textbf{8} (2014).

\bibitem{Murakami2014}
H. Murakami, K. Serita, Y. Maekawa, S. Fujiwara, E. Matsuda, S. Kim, I. Kawayama and M. Tonouchi \enquote{Scanning laser THz imaging system} J. Phys. D: Appl. Phys. \textbf{47} (2014).

\bibitem{Giordano2018}
M.C. Giordano, S. Mastel, C. Liewald, L.L. Columbo, M. Brambilla, L. Viti, A. Politano, K. Zhang, L. Li, A.G. Davies, E.H. Linfield, R. Hillenbrand, F. Keilmann, G. Scamarcio, and M.S. Vitiello \enquote{Phase-resolved terahertz self-detection near-field microscopy} Opt. Express \textbf{26} (2018).

\bibitem{Thomas2022}
L. Thomas, T. Hannotte, C. N. Santos, B. Walter, M. Lavancier, S. Eliet, M. Faucher, J.-F. Lampin, and R. Peretti \enquote{Imaging of THz Photonic Modes by Scattering Scanning Near-Field Optical Microscopy} ACS Applied Materials and Interfaces \textbf{14} (2022).

\bibitem{Bai2020}
Z. Bai, Y. Liu, R. Kong, T. Nie, Y. Sun, H. Li, T. Sun, C. Pandey, Y. Wang, H. Zhang, Q. Song, G. Liu, M. Kraft, W. Zhao, X. Wu, and L. Wen \enquote{Near-field Terahertz Sensing of HeLa Cells and Pseudomonas Based on Monolithic Integrated Metamaterials with a Spintronic Terahertz Emitter} ACS Appl. Mater. Interfaces \textbf{12} (2020).

\bibitem{Chen2020}
S.-C. Chen, Z. Feng, J. Li, W. Tan, L.-H. Du, J. Cai, Y. Ma, K. He, H. Ding, Z.-H. Zhai, Z.-R. Li, C.-W. Qiu, X.-C. Zhang, and L.-G. Zhu \enquote{Ghost spintronic THz-emitter-array microscope} Light: Science and Applications \textbf{9} (2020).
\end{thebibliography}
\end{document}